# Satellite motion in a non-singular gravitational potential

Ioannis Haranas[1] and Spiros Pagiatakis[1,2]

[1] *Dept. of Physics and Astronomy, York University, 4700 Keele Street, Toronto, Ontario, M3J 1P3, Canada*
e mail: ioannis@yorku.ca

[2] *Dept. of Earth & Space Science & Engineering, York University, 4700 Keele Street, Toronto, Ontario, M3J 1P3, Canada*
e mail: spiros@yorku.ca

**Abstract**
We study the effects of a non-singular gravitational potential on satellite orbits by deriving the corresponding time rates of change of its orbital elements. This is achieved by expanding the non-singular potential into power series up to second order. This series contains three terms, the first been the Newtonian potential and the other two, here $R_1$ (first order term) and $R_2$ (second order term), express deviations of the singular potential from the Newtonian. These deviations from the Newtonian potential are taken as disturbing potential terms in the Lagrange planetary equations that provide the time rates of change of the orbital elements of a satellite in a non-singular gravitational field. We split these effects into secular, low and high frequency components and we evaluate them numerically using the low Earth orbiting mission Gravity Recovery and Climate Experiment (GRACE). We show that the secular effect of the second-order disturbing term $R_2$ on the perigee and the mean anomaly are $4''.307 \times 10^{-9}$/a , and $-2''.533 \times 10^{-15}$/a, respectively. These effects are far too small and most likely cannot easily be observed with today's technology. Numerical evaluation of the low and high frequency effects of the disturbing term $R_2$ on low Earth orbiters like GRACE are very small and undetectable by current observational means.

## 1. Introduction

A non-singular gravitational potential may take the following form [Williams, 2001]

$$V(r) = -\frac{GM_p}{r} e^{-\lambda/r}, \tag{1}$$

where constant $\lambda$ is defined as follows

$$\lambda = \frac{GM_p}{c^2}, \tag{2}$$

$G$ is Newton's gravitational constant, $M_p$ is the mass of the planetary body that produces the potential, $c$ is the speed of light, and $r$ is the radial distance of the satellite from the center of mass of the planetary body.

The goal of this contribution is to examine the possibility of validating this non-singular potential by studying satellite orbit perturbations that might result from the deviation of this singular potential from the Newtonian one. Various satellite effects can conveniently be expressed as orbital element time rates of change, which are observable by modern geodetic techniques. In general, the well-known Lagrange planetary equations, as they are presented for instance in Kaula [2000], link the orbital element time derivatives to their cause, a disturbing (or perturbing) potential. Here, disturbing potential implies any deviation of the total potential from a central Newtonian field. Accepting that Eq. (1) holds true, we can write $V(r)$ as a central Newtonian potential plus other terms that constitute the disturbing components. These disturbing components can then be entered separately into the Lagrange planetary equations to study their effects on the satellite central field (Keplerian) orbit, with the hope that we can see some measurable orbital element time rates of change and thus observationally verify or disprove Eq. (1).

The Lagrange planetary equations contain the derivatives of an appropriate disturbing potential $R$ with respect to the orbital elements. Following Kaula [2000] we can write the Lagrange planetary equations as follows:

$$\frac{da}{dt} = \frac{2}{na} \frac{\partial R}{\partial M}, \tag{3}$$

$$\frac{de}{dt} = \frac{(1-e^2)}{nea^2} \frac{\partial R}{\partial M} - \frac{\sqrt{1-e^2}}{nea^2} \frac{\partial R}{\partial \omega}, \tag{4}$$

$$\frac{d\omega}{dt} = \frac{\sqrt{1-e^2}}{na^2 e} \frac{\partial R}{\partial e} - \frac{\cos i}{na^2 \sqrt{1-e^2} \sin i} \frac{\partial R}{\partial i}, \tag{5}$$

$$\frac{di}{dt} = \frac{\cos i}{na^2 \sqrt{1-e^2} \sin i} \frac{\partial R}{\partial \omega} - \frac{1}{na^2 \sqrt{1-e^2} \sin i} \frac{\partial R}{\partial \Omega}, \tag{6}$$

$$\frac{d\Omega}{dt} = \frac{1}{na^2 \sqrt{1-e^2} \sin i} \frac{\partial R}{\partial i}, \tag{7}$$

$$\frac{dM}{dt} = n - \frac{(1-e^2)}{nea^2} \frac{\partial R}{\partial e} - \frac{2}{na} \frac{\partial R}{\partial a}, \tag{8}$$

where $a$ is the orbital semimajor axis, $e$ is the orbital eccentricity, $\omega$ is the argument of the perigee, $i$ is the orbital inclination, $\Omega$ is the argument of the ascending node, $M$ is the mean anomaly[1], and $n = (GM_p / a^3)^{1/2}$ is the mean motion of the satellite. In our study, the disturbing potential $R = R(a, e, \omega, i, \Omega, M)$ contains only the deviations of the non-singular potential (cf. Eq. (1)) from the central Newtonian. To obtain $R$, we use power series expansion of Eq. (1) that allows expressing the non-singular potential as the sum of a central potential and its disturbing terms. The disturbing terms form $R$, which after appropriate transformations, it can be written as a function of the orbital elements and eccentricity functions [e.g. Kaula, 2000] so that its derivatives with respect to the orbital elements, as required by the planetary equations, can be taken (cf. Eqs. (3)-(8)). Using the Lagrange planetary equations we can then numerically evaluate the time rates of change of the orbital elements due to $R$ and thus, we will be able to assess whether the non-singular potential can or cannot be verified experimentally.

## 2. The disturbing potential

Without loss of generality, we consider a satellite that orbits the Earth at a certain radial distance $r$ from the geocenter under the influence of the non-singular potential given by Eq. (1). We can expand the exponential term of Eq. (1) into power series and keeping terms up to second degree we obtain

$$V(r) \cong -\frac{GM_p}{r} + \frac{GM_p \lambda}{r^2} - \frac{GM_p \lambda^2}{2r^3}. \tag{9}$$

Using Eq. (2) we can write Eq. (9) as follows

$$V(r) \cong V_N + \frac{G^2 M_p^2}{c^2 r^2} - \frac{G^3 M_p^3}{2c^4 r^3} \cong V_N + R_1 + R_2, \tag{10}$$

where $V_N$ is the central Newtonian potential, and the other two terms in the RHS of (10) express deviations of the non-singular potential from the central Newtonian and are denoted as disturbing components $R_1$ and $R_2$, respectively. The third term in the RHS of (10) is inversely proportional to $r^3$, with a similar radial dependence to the general relativistic potential that reads $V_{GR} = GMh^2/c^2 r^3$ [Murray and Dermott, 1999], where $h = \sqrt{GMa(1-e^2)}$ is the angular momentum per unit mass of the primary body. In particular, using the relativistic potential and substituting for the angular momentum $h$ we can write $R_2$ as a function of the relativistic potential as follows

$$R_2 = \frac{GM_p}{2c^2 a(1-e^2)} V_{GR}. \tag{11}$$

---

[1] For a more detailed definition of the orbital elements see Vallado [2007].

Theory predicts that the relativistic potential causes secular perigee/perihelion variations in the orbit of a satellite (natural or artificial) orbiting a massive body [Glosh, 2000]. The term $1/r^{\ell+1}$, where $\ell$ is integer (see below), can be written as a function of the eccentricity functions $G_{\ell pq}(e)$ and satellite orbital elements in the apparent right ascension system as follows [Kaula, 2000; p.35]:

$$\frac{1}{r^{\ell+1}}\binom{\cos}{\sin}[(\ell-2p)(\omega+f)+m(\Omega-\Theta)]=\frac{1}{a^{\ell+1}}\sum_{q=-\infty}^{+\infty}G_{\ell pq}(e)\binom{\cos}{\sin}[(\ell-2p)\omega+(\ell-2p+q)M+m(\Omega-\Theta)], \quad (12)$$

where $f$ is the true anomaly, $\Theta$ is the Greenwich sidereal time, $\ell$ is the degree and $m$ is the order of the spherical harmonic expansion of the potential, $(p, q) \in Z$ and $0 \le p \le \ell$. The indices $\ell, p, q, m$ identify the eccentricity function and also the trigonometric argument associated with a particular spherical harmonic term of degree $\ell$ and order $m$. These terms arise from the potential of the Earth when it is expressed in terms of spherical harmonics as given in Kaula [cf. Eq. (1.31); Kaula, 2000]. Using Eq. (12), we write the two terms $R_1$ and $R_2$ as functions of the orbital elements in the following way:

$$R_1 = \frac{G^2 M_p^2}{c^2}\left[\frac{1}{a^{\ell+1}}\sum_{q=-\infty}^{+\infty}G_{\ell pq}(e)\binom{\cos}{\sin}[(\ell-2p)\omega+(\ell-2p+q)M+m(\Omega-\Theta)]\right], \quad (13)$$

$$R_2 = \frac{G^3 M_p^3}{2c^4}\left[\frac{1}{a^{\ell+1}}\sum_{q=-\infty}^{+\infty}G_{\ell pq}(e)\binom{\cos}{\sin}[(\ell-2p)\omega+(\ell-2p+q)M+m(\Omega-\Theta)]\right]. \quad (14)$$

### 3. The secular disturbing potentials and time rates of change of the orbital elements

Next, we examine only the secular terms resulting from $R_1$ and $R_2$ in Eqs. (13) and (14), respectively. We can do this by eliminating the low frequency term $\omega$ from Eqs. (13) and (14) by setting $\ell-2p=0$. Similarly, from Eqs. (13) and (14), we eliminate the terms that are varying with high frequency, i.e., the terms that are functions of the mean anomaly $M$, and $(\Omega-\Theta)$. This can be achieved by setting their respective coefficients to zero, which results in $(\ell-2p+q)=0$, and $m = 0$, which imply $q = 0$ since $\ell = 2p$. These conditions must hold simultaneously and finally, Eqs. (13) and (14) become:

$$R_{1_S} = \frac{G^2 M_p^2}{c^2 a^2} G_{1,1/2,0}, \quad (15)$$

$$R_{2_S} = \frac{G^3 M_p^3}{2c^4 a^3} G_{2,1,0}, \quad (16)$$

where subscript "S" signifies "secular." Clearly, in the case of $R_1$, we have $\ell = 1$ (cf. Eq. (12)) which implies $p=1/2$. However, $p$ must always be an integer [Vallado, 2007] and in addition $\ell = 1$ harmonic is identically zero because the coordinate system is geocentric. This indicates that $R_1$ is not physically meaningful and thus disregarded from further consideration. In order to proceed with the calculation of the secular time rates of change of the orbital elements due to $R_2$, we substitute Eq. (16) into the Lagrange planetary equations. The calculation of eccentricity function $G_{\ell pq}(e)$ is not a trivial process because it requires the use of the so called Hansen coefficients $X_k^{n,m}$. Following Giacaglia, [1976] we have that

$$G_{\ell pq}(e) = X_{\ell-2p+q}^{-(\ell+1),(\ell-2p)}, \quad (17)$$

and the corresponding eccentricity function $G_{2,1,0}(e)$ becomes [Kaula, 2000]

$$G_{2,1,0}(e) = (1-e^2)^{-3/2}. \quad (18)$$

To demonstrate the relation/difference between the general relativity and the non-singular potential effects on the above two orbital elements derived herein, we consider the following expressions for the prediction of the secular rates of change of the perigee [Lucchesi, 2003] and mean anomaly, respectively [Schwarzschild, 1916].

$$\left(\frac{d\omega}{dt}\right)_{GR_S} = \frac{3(GM)^{3/2}}{c^2(1-e^2)a^{5/2}}, \tag{19}$$

$$\left(\frac{dM}{dt}\right)_{GR_S} = \frac{3(GM_p)^{3/2}}{c^2 a^{5/2}(1-e^2)^{1/2}}. \tag{20}$$

Using $R_{2_S}$ we obtain

$$\left(\frac{d\omega}{dt}\right)_{R_{2_S}} = \frac{1}{6n}\left(\frac{d\omega}{dt}\right)^2_{GR_S}, \tag{21}$$

$$\left(\frac{dM}{dt}\right)_{R_{2_S}} = n + \frac{1}{6n}\left[1 - \frac{1}{(1-e^2)^{7/2}}\right]\left[\left(\frac{d\omega}{dt}\right)\left(\frac{dM}{dt}\right)\right]_{GR_S} \tag{22}$$

where the subscript $R_{2_S}$ signifies secular changes caused by $R_2$.

### 4. The low frequency disturbing potential and the time rates of change of the orbital elements

Focusing on the low frequency terms of $R_2$, we eliminate the terms from Eq. (14) that vary with high frequency. This can be achieved by setting their respective coefficients to zero resulting to $\ell - 2p + q = 0$ and $m = 0$. For the $1/r^3$ term in (12) we have that $\ell = 2$ and $0 \le p \le \ell$ and $q = 2p - \ell$, which implies that $q \in [-2,\ 2]$ therefore, Eq. (14) becomes

$$R_{2_{LS}} = \frac{G^3 M_p^3}{2c^4}\left[\frac{1}{a^3}\sum_{q=-2}^{+2} G_{2,p,q}(e)\cos[(2-2p)\omega]\right]. \tag{23}$$

where subscript $LS$ indicates "Low" frequency components. Substituting Eq. (23) in the Lagrange planetary equations we obtain the following equations for the low frequency time rates of change of the orbital elements due to $R_2$, and the corresponding sine terms will be zero. Therefore, the only non-zero time rates are

$$\frac{de}{dt} = \frac{G^3 M_p^3 \sqrt{1-e^2}}{2nc^4 e a^5}\sum_{p=0}^{2}(2-2p)G_{2,p,2p-2}(e)\sin[(2-2p)\omega], \tag{24}$$

$$\frac{d\omega}{dt} = \frac{G^3 M_p^3 \sqrt{1-e^2}}{2nec^4 a^5}\sum_{p=0}^{2}\frac{\partial G_{2,p,2p-2}(e)}{\partial e}\cos[(2-2p)\omega], \tag{25}$$

$$\frac{di}{dt} = -\frac{\cot i\, G^3 M_p^3}{2nc^4 a^5 \sqrt{1-e^2}}\sum_{p=0}^{2}(2-2p)G_{2,p,2p-2}(e)\sin[(2-2p)\omega], \tag{26}$$

$$\frac{dM}{dt} = n - \frac{(1-e^2)G^3 M_p^3}{nec^4 a^5}\sum_{p=0}^{2}\frac{\partial G_{2,p,2p-2}(e)}{\partial e}\cos[(2-2p)] + \frac{3G^3 M^3}{nec^4 a^5}\sum_{p=0}^{2}G_{2,p,2p-2}(e)\cos[(2-2p)\omega]. \tag{27}$$

Carrying out the summation in the above equations we obtain

$$\frac{de}{dt} = \frac{G^3 M_p^3 \sqrt{1-e^2}}{2nc^4 e a^5}\left[2(G_{2,0,-2}(e) + G_{2,2,2}(e))\sin(2\omega)\right] = 0, \tag{28}$$

$$\frac{d\omega}{dt} = \frac{G^3 M_p^3 \sqrt{1-e^2}}{2nec^4 a^5}\left[\frac{\partial G_{2,1,0}(e)}{\partial e} + \left(\frac{\partial G_{2,0,-2}(e)}{\partial e} + \frac{\partial G_{2,2,2}}{\partial e}\right)\cos(2\omega)\right] = \frac{3G^3 M^3}{2c^4 a^5(1-e^2)^2}, \tag{29}$$

$$\frac{di}{dt} = -\frac{\cot i\, G^3 M_p^3}{2nc^4 a^5 \sqrt{1-e^2}}\left[2(G_{2,0,-2}(e) + G_{2,2,2}(e))\sin(2\omega)\right] = 0, \tag{30}$$

$$\frac{dM}{dt} = n - \frac{(1-e^2)G^3 M_p^3}{nec^4 a^5}\left[\left(\frac{\partial G_{2,0,-2}(e)}{\partial e} + \frac{\partial G_{2,2,2}(e)}{\partial e}\right)\cos(2\omega)\right] + \frac{3G^3 M^3}{c^4 nea^5}\left[2(G_{2,0,-2}(e) + G_{2,2,2})\cos(2\omega)\right] = n. \tag{31}$$

Using the tabulated expressions of the eccentricity functions we have that [Kaula, 2000]

$$G_{2,0,-2}(e) = G_{2,0,2}(e) = 0, \tag{32}$$

and using Eq. (18),

$$\frac{\partial G_{2,1,0}}{\partial e} = 3e(1-e^2)^{-5/2}. \tag{33}$$

Eqs. (28)-(31) above give the low frequency variations of the orbital elements due to $R_2$ from which the non-zero rates of change can be written as follows

$$\frac{d\omega}{dt} = \frac{3G^3 M_p^3}{2nc^4 a^5 (1-e^2)^2}, \tag{34}$$

$$\frac{dM}{dt} = n. \tag{35}$$

We see that from all the orbital elements, only the argument of the perigee is affected by the low frequency term due to $R_2$. This is a fraction of the secular variation given by general relativity calculated using Eq. (23) and therefore, we have that

$$\dot{\omega}_{R_{2LS}} = \frac{1}{6} \dot{\omega}_{GR_S}^2. \tag{36}$$

## 5. High-frequency disturbing potentials and time rates of change of the orbital elements

In order to obtain the high frequency components of the disturbing function $R_2$, we simply eliminate the low-frequency terms in (14) and we get

$$R_{2_H} = \frac{G^3 M_p^3}{2c^4} \left[ \frac{1}{a^3} \sum_{m=0}^{2} \sum_{q=1}^{\infty} G_{2,1,q}(e) \cos(qM + m(\Omega - \Theta)) \right], \tag{37}$$

where subscript "$H$" signifies "high frequency." Substituting Eq. (37) in the Lagrange equations we obtain the following high frequency variations of the orbital elements. We proceed with the derivation of the high frequency effects arising from $R_2$ by summing over index $q \leq 4$ for, when $q>4$, the effects of $R_2$ are O($10^{-18}$) on $\dot{a}$, O($10^{-21}$) on $\dot{e}$, O($10^{-13}$) on $\dot{\omega}$, and O($10^{-20}$) on $\dot{i}$ and O($10^{-13}$) on $\dot{M}$. For the $R_2$ disturbing term we obtain the following non-zero time rates of change

$$\frac{da}{dt} = -\frac{G^3 M_p^3}{nc^4 a^4} \sum_{m=0}^{2} \sum_{q=1}^{\infty} qG_{2,1,q}(e) \sin[qM + m(\Omega - \Theta)], \tag{38}$$

$$\frac{de}{dt} = -\frac{(1-e^2)G^3 M_p^3}{2nec^4 a^5} \sum_{m=0}^{2} \sum_{q=1}^{\infty} qG_{2,1,q}(e) \sin[qM + m(\Omega - \Theta)], \tag{39}$$

$$\frac{d\omega}{dt} = \frac{G^3 M_p^3 \sqrt{1-e^2}}{2nec^4 a^5} \sum_{m=0}^{2} \sum_{q=1}^{\infty} \frac{\partial G_{2,1,q}(e)}{\partial e} \cos[qM + m(\Omega - \Theta)], \tag{40}$$

$$\frac{di}{dt} = \frac{G^3 M_p^3}{nc^4 a^5 \sin i \sqrt{1-e^2}} \sum_{m=0}^{2} \sum_{q=1}^{\infty} mG_{2,1,q}(e) \sin[qM + m(\Omega - \Theta)], \tag{41}$$

$$\frac{dM}{dt} = n + \frac{3G^3 M_p^3}{nc^4 a^5} \left( \sum_{m=0}^{2} \sum_{q=1}^{\infty} \left( G_{2,1,q}(e) - \frac{(1-e^2)}{6e} \frac{\partial G_{2,1,q}(e)}{\partial e} \right) \cos[qM + m(\Omega - \Theta)] \right). \tag{42}$$

In the above equations we also need the following eccentricity functions ($q \leq 4$).

$$G_{2,1,1}(e) = \frac{3e}{2} + \frac{27e^3}{16} \tag{43}$$

$$G_{2,1,2}(e) = \frac{9e^2}{4} + \frac{7e^4}{4} \tag{44}$$

$$G_{2,1,3}(e) = \frac{53e^3}{16} - \frac{393e^5}{256} \tag{45}$$

$$G_{2,1,4}(e) = \frac{77\,e^4}{16} - \frac{129\,e^6}{160} \tag{46}$$

Substituting Eqs. (43) – (46) into Eqs. (38) – (42) we get

$$\frac{da}{dt} = -\frac{G^3 M_p^3}{nc^4 a^4} \left[ \begin{array}{l} \left[ 2\left(\frac{3e}{2} + \frac{27e^3}{16}\right)\sin\left(\frac{1}{2}(2M - \Theta + \Omega)\right) + 4\left(\frac{9e^2}{4} + \frac{7e^4}{4}\right)\sin\left(\frac{1}{2}(4M - \Theta + \Omega)\right) \\ + \frac{30e^2}{16}\sin\left(\frac{1}{2}(6M - \Theta + \Omega)\right) + \frac{154e^4}{4}\sin\left(\frac{1}{2}(8M - \Theta + \Omega)\right) \end{array} \right]\cos\left(\frac{1}{2}(\Theta - \Omega)\right) \\ + \left(\frac{3e}{2} + \frac{27e^3}{16}\right)\sin(M + 2(\Omega - \Theta)) + 2\left(\frac{9e^2}{4} + \frac{7e^4}{4}\right)\sin(2M + 2(\Omega - \Theta)) \\ + \frac{15e^3}{16}\sin(3M + 2(\Omega - \Theta)) + \frac{77e^4}{4}\sin(4M + 2(\Omega - \Theta)) \end{array} \right]$$

(47)

$$\frac{de}{dt} = -\frac{G^3 M_p^3 (1-e^2)}{2enc^4 a^5} \left[ \begin{array}{l} \left[ 2\left(\frac{3e}{2} + \frac{27e^3}{16}\right)\sin\left(\frac{1}{2}(2M - \Theta + \Omega)\right) \\ + 4\left(\frac{9e^2}{4} + \frac{7e^4}{4}\right)\sin\left(\frac{1}{2}(4M - \Theta + \Omega)\right) + \frac{30e^3}{16}\sin\left(\frac{1}{2}(6M - \Theta + \Omega)\right) \\ + \frac{154e^4}{4}\sin\left(\frac{1}{2}(8M - \Theta + \Omega)\right) + \end{array} \right]\cos\left(\frac{1}{2}(\Theta - \Omega)\right) \\ \left(\frac{3e}{2} + \frac{27e^3}{16}\right)\sin(M + 2(\Omega - \Theta)) \\ + 2\left(\frac{9e^2}{4} + \frac{7e^4}{4}\right)\sin(2M + 2(\Omega - \Theta)) + \frac{15e^3}{16}\sin(3M + 2(\Omega - \Theta)) \\ + \frac{77e^4}{4}\sin(4M + 2(\Omega - \Theta)) \end{array} \right]$$

(48)

[2] $G_{2,1,1}(e)$ and $G_{2,1,2}(e)$ evaluated herein are identical to those given in Kaula [2000] and Vallado [2007].

$$\frac{d\omega}{dt} = -\frac{G^3 M_p^3 \sqrt{1-e^2}}{nec^4 a^5} \begin{bmatrix} \left[ 2\left(\frac{3}{2} + \frac{81e^2}{16}\right)\sin\left(\frac{1}{2}(2M - \Theta + \Omega)\right) \\ + 2\left(\frac{9e}{2} + 7e^3\right)\sin\left(\frac{1}{2}(4M - \Theta + \Omega)\right) + 2\left(\frac{159e^2}{16} + \frac{1965e^4}{256}\right)\sin\left(\frac{1}{2}(6M - \Theta + \Omega)\right) \right]\cos\left(\frac{1}{2}(\Theta - \Omega)\right) \\ + 2\left(\frac{77e^3}{4} + \frac{387e^5}{80}\right)\sin\left(\frac{1}{2}(8M - \Theta + \Omega)\right) \\ + \left(\frac{3}{2} + \frac{81e^2}{16}\right)\cos(M + 2(\Omega - \Theta)) \\ + \left(\frac{9e}{2} + 7e^3\right)\cos(2M + 2(\Omega - \Theta)) \\ + \left(\frac{159e^2}{16} + \frac{1965e^4}{256}\right)\cos(3M + 2(\Omega - \Theta)) + \left(\frac{77e^3}{4} + \frac{387e^5}{80}\right)\cos(4M + 2(\Omega - \Theta)) \end{bmatrix} \tag{49}$$

$$\frac{di}{dt} = \frac{G^3 M_p^3 \csc i}{nc^4 a^5 \sqrt{1-e^2}} \begin{bmatrix} \left(\frac{3e}{2} + \frac{27e^3}{16}\right)\cos(M - \Theta + \Omega) + \left(\frac{9e^2}{4} + \frac{7e^4}{4}\right)\cos(2M - \Theta + \Omega) \\ + \frac{5e^3}{16}\cos(3M - \Theta + \Omega) + \frac{77e^4}{16}\cos(4M - \Theta + \Omega) \end{bmatrix} \tag{50}$$

$$\frac{dM}{dt} = n - \frac{G^3 M_p^3}{nc^4 a^3} \begin{bmatrix} 3 \begin{bmatrix} 2\left(\frac{3e}{2} + \frac{27e^3}{16} - \frac{(1-e^2)}{6e}\left(\frac{3}{2} + \frac{81e^2}{16}\right)\right)\sin\left(\frac{1}{2}(2M - \Theta + \Omega)\right) \\ + 2\left(\frac{9e^2}{4} + \frac{7e^4}{4} - \frac{(1-e^2)}{6e}\left(\frac{9e}{2} + 7e^3\right)\right)\sin\left(\frac{1}{2}(4M - \Theta + \Omega)\right) \\ + 2\left(\frac{53e^3}{16} + \frac{393e^5}{256} - \frac{(1-e^2)}{6e}\left(\frac{159e^2}{16} + \frac{1965e^4}{256}\right)\right)\sin\left(\frac{1}{2}(6M - \Theta + \Omega)\right) \\ 2\left(\frac{77e^4}{16} + \frac{129e^6}{160} - \frac{(1-e^2)}{6e}\left(\frac{77e^3}{4} + \frac{387e^5}{80}\right)\right)\sin\left(\frac{1}{2}(8M - \Theta + \Omega)\right) \end{bmatrix} \cos\left(\frac{1}{2}(\Theta - \Omega)\right) \end{bmatrix}. \tag{51}$$

## 6. Numerical Results

We calculate the secular orbital element changes (cf. Eqs. (15)-(16)) specifically for the *Gravity Recovery and Climate Experiment – GRACE* mission, using the orbital parameters of GRACE-A satellite that has $a$= 6876.4816 km, and $e$ = 0.00040989, and therefore $n$ = 0.001100118 rad/s = 15.113 rev/d, $i$ = 89.025446°, $\omega$ = 302.414244°, $\Omega$ = 354.447149°, $M$ = 80.713591° [*http://www.csr.utexas.edu/grace/newsletter/archive/august2002.html*]. Because all derived Eqs. (19)-(22) are inversely proportional to different powers of the semimajor axis, the secular rates of change of the orbital elements due to general relativity diminish rapidly for higher altitude satellites thus, the choice of GRACE mission (low orbit). Substituting these values in Eqs. (23) and (24) we obtain the corresponding secular general relativistic effects on $\omega$ and $M$ as follows

$$\left(\frac{d\omega}{dt}\right)_{GR_s} = 13''.30/\text{a}, \tag{53}$$

$$\left(\frac{dM}{dt}\right)_{GR_s} = 13''.30/a. \tag{54}$$

Similarly, using Eqs. (21) and (22) we calculate the corresponding secular rates of change of $\omega$ and $M$ due to $R_2$ for which we obtain

$$\left(\frac{d\omega}{dt}\right)_{R_{2S}} = 4''.307 \times 10^{-9} /a \tag{55}$$

$$\left(\frac{dM}{dt}\right)_{R_{2S}} = -2''.533 \times 10^{-15} /a. \tag{56}$$

The low frequency maximum effect on the perigee is $9''.525 \times 10^{-17}$ and far too small, to be observed with today's technology.

Finally, for the numerical calculation of the high frequency effects of $R_{2H}$ on the orbital element time rates of change we choose to calculate only the maximum effect because Eqs (47)-(51) contain many sine waves of various frequencies. This can be done by setting all trigonometric terms equal to unity implying that all constituent waves are in phase. The maximum effects on $a$, $e$, $\omega$, $i$, and $M$ are $-5.865 \times 10^{-18}$ m, $-1.040 \times 10^{-21}$, $5''.240 \times 10^{-13}$, $5''.864 \times 10^{-20}$, and $-3''.492 \times 10^{-13}$, respectively, whereas the effect on $\Omega$ is zero. Apparently, these maximum variations are far too small to be observed with today's technology.

Next, we calculate the secular effects of $R_2$ and, we find that the corresponding time rates of change of the perigee and mean anomaly are extremely small, namely $\dot{\omega}_{R_{2S}} = 4''.307 \times 10^{-9}/a$, and $\dot{M}_{R_{2S}} = -2''.553 \times 10^{-15}/a$, leaving the time rate of change of the mean anomaly practically unchanged, and equal to that of the Newtonian field. With reference to GRACE-A satellite only, these rates of change of the perigee are by far smaller than any technology can measure today, and require very long orbiting times that far exceed the design lifetime of low Earth orbiters. For natural satellites or planets like Mercury that is the closest planet to the massive Sun, there might actually be a possibility to obtain measurable effects. For Mercury with a semimajor axis $a = 57.91 \times 10^6$ km and eccentricity $e = 0.205631752$ [Vallado, 2007] we obtain $\dot{\omega}_{R_{2s}} = 4''.00 \times 10^{-9}/a$, and $\dot{M}_{R_{2s}} = -6''.546 \times 10^{-13}/a$, still much too small to accumulate to a measurable effect in time-scales of centuries in a way similar to the relativistic effect of the perihelion of Mercury.

## 7. Conclusions
We used Kaula's approach to transform and validate the non-singular potential given by Eq.(1) using satellite orbit perturbations. Examining the high frequency terms we found that their corresponding effects are far too small to be detected. Similarly, we found that the low frequency effect of $R_2$ on the perigee is far too small to be observed with today's technology. In addition, and for GRACE mission, the calculated secular changes related to $R_2$ were found to be extremely small, and impossible to observe with current technology. In conclusion, Eq. (1) cannot be verified using low Earth orbiters, at least with the current technology.

**Acknowledgements:** This research was financially supported by the Natural Sciences and Engineering Research Council (NSERC) of Canada. We thank the two anonymous reviewers for their thoughtful comments and suggestions that significantly improved the original manuscript.